\documentclass[aps,reprint,groupedaddress,superscriptaddress,longbibliography]{revtex4-2}
\usepackage{amsmath}
\usepackage{amssymb}
\usepackage{amsfonts}
\usepackage[pdftex]{graphicx,color}
\usepackage{textgreek}
\usepackage{makecell,tabularx}
\setcellgapes{3pt}

\begin{document}

\title{Undulating compression and multi-stage relaxation in a granular column consisting of dust particles or glass beads}

\author{Felipe Pacheco-V\'azquez}
\affiliation{Instituto de F\'isica, Benem\'erita Universidad Aut\'onoma de Puebla, Apartado Postal J-48, Puebla 72570, Mexico}
\affiliation{Department of Earth and Environmental Sciences, Nagoya University, Furocho, Chikusa, Nagoya 464-8601, Japan}

\author{Tomomi Omura}
\affiliation{Institute of Education Center of Advanced Education, Osaka Sangyo University, 3-1-1 Nakagaito, Daito-shi, Osaka 574-8530, Japan}
\affiliation{Department of Earth and Environmental Sciences, Nagoya University, Furocho, Chikusa, Nagoya 464-8601, Japan}

\author{Hiroaki Katsuragi}
\affiliation{Department of Earth and Space Science, Osaka University, 1-1 Machikaneyama, Toyonaka 560-0043, Japan}
\affiliation{Department of Earth and Environmental Sciences, Nagoya University, Furocho, Chikusa, Nagoya 464-8601, Japan}

\date{\today}

\begin{abstract}
  For fundamentally characterizing the effect of hierarchical structure in granular matter, a set of compression-relaxation tests for dust particles and glass beads confined in a cylindrical cell was performed. Typical diameter of both grains is approximately 1~mm. However, dust particles are produced by binding tiny ($\sim 5$~{\textmu}m) glass beads. The granular columns were compressed with a piston until reaching a maximum load force of 20~N with a constant compression rate $v$ ($0.17 \leq v \leq 2000$~{\textmu}m~s$^{-1}$). After that, the piston was stopped and the relaxation process was quantified. From the experimental results, we found that the compression force $F$ nonlinearly increases with the increase of compression stroke $z$ depending on particles. Besides, periodic undulation and sudden force drops were observed on $F(z)$ in dust particles and glass beads, respectively. The relaxation process was characterized by an exponential decay of stress followed by a logarithmic dependence one in both kinds of particles. These experimental findings are the main point in this study. To understand the underlying physics governing the compression mechanics, we assumed empirical forms of $F(z)$; $F\propto z^{\alpha}$ for dust particles and $F \propto \exp(z/z_G)$ for glass beads ($\alpha=2.4$ and $z_G=70$~{\textmu}m). Then, we found that the growing manners of periodic undulation and force drops were identical to those of mean compression forces, i.e., power law in dust particles and exponential in glass beads. In addition, the undulation amplitude and wavelength decreased as $v$ increased in dust-particles compression. On the basis of experimental results and the difference between dust particles and glass beads, we also discuss the origin of undulation and the physical meaning of granular-compression models used in engineering fields. 
\end{abstract}


\maketitle

\section{Introduction}
Ubiquity and importance of granular matter in various natural phenomena have been recognized by recent numerous studies on granular behaviors~\cite{Andreotti:2013}. Granular matter has been usually defined by a collection of rigid dissipative particles. The principal sources of dissipation are inelastic collisions and friction. While this type of typical granular matter exhibits various counterintuitive behaviors, a different type of granular matter, hierarchical granular matter, has also been studied recently. 

Hierarchical granular matter consists of granules of tiny particles. When we treat tiny particles, various cohesive forces (e.g., capillary force, electrostatic force, van der Waals force) work as a binder to form granules. By collecting the formed granules, hierarchical granular matter is comprised. Hierarchical granular matter is characterized by two distinct length scales. One is the length scale of tiny monomers that construct porous dust particles. The other length scale is the size of dust particles. Due to this complexity, behaviors of hierarchical granular matter could become quite different from conventional granular matter. In general, understanding of the hierarchical structure is a crucial key to reveal the complex nature of various phenomena. Therefore, we focus on the role of the hierarchical structure in the physics of granular matter.

Large dust particles, usually called dust aggregates, have attracted planetary physicists because they are plausible candidates for the material of planetesimal formation~\cite{Blum:2008,Blum:2018}. To reveal the physical properties, mechanical characteristics on dust aggregates have been studied~\cite{Sirono:2004,Blum:2006,Kataoka:2013,Omura:2018,Katsuragi:2017}. Recently, the existence of such porous particles/aggregates on small celestial bodies (comets/asteroids) has been reported by planetary explorations~\cite{Bentley:2016,Okada:2020}. If comets consist of hierarchical dust particles, mechanical and thermal properties must be influenced by the hierarchical structure~\cite{Skorov:2012,Arakawa:2020}. Furthermore, a granular porous projectile impacting onto a granular target produces particular crater shapes as experimentally demonstrated in~\cite{PachecoVazquez:2011}. Undoubtedly, the fundamental understanding of dust particles/aggregates is crucial to elucidate the planetary-scale phenomena as well as laboratory-scale phenomena. 

The impact response of the collection of dust aggregates (hierarchical granular matter) was also studied recently~\cite{Katsuragi:2018}. In the study, impact-induced expansion of granular cluster consisting of dust particles or glass beads was measured and the similarity between dust particles and glass beads was reported. However, whether this similarity is truly universal or not should be checked by other experimental conditions. By revealing fundamental features of hierarchical granular matter through the systematic investigation, we might be able to establish a novel type of soft matter. Actually, one can easily find various entities of hierarchical granular matter even in our everyday life, e.g., snow balls, granules of sugar, medicines, etc. Thus, the fundamental study of hierarchical granular matter affects various kinds of science and industries. 

As a basic methodology, uniaxial compression test of granular matter has been frequently performed both in soil mechanics~\cite{Knapett:2012} and powder engineering~\cite{Denny:2002}. Because these fields have been independently developed, different models have been applied to the analysis of similar compression data. In soil mechanics field, compression index has been used to characterize the pressure-dependent void shrinkage in granular (soil) material~\cite{Knapett:2012}. In powder engineering field, on the other hand, the compressibility of void space in granular matter has been characterized by using bulk modulus~\cite{Denny:2002}. While these quantities have been independently used in each research field, their similarity and difference have not been sufficiently analyzed so far. Fundamental physical consideration is necessary to properly compare and unify these models. Of course, uniaxial compression test is one of the most fundamental methods to characterize soft materials. To characterize mechanical properties of novel soft materials such as hierarchical granular matter, uniaxial compression tests should be performed at first. 

In the case of brittle granular materials, propagation of upward compaction bands was reported during vertical (uniaxial) compression of snow~\cite{Barraclough:2016} and cereals~\cite{Valdes:2011}. Numerical simulations are able to reproduce this propagating band and have been helpful to determine the conditions required to observe such dynamics~\cite{Guillard:2015}. This propagating band is characterized by sudden stress drops produced by the breakage of particles at the bottom of the bed that allows its relaxation and further rearrangement.

In this study, we carried out a set of simple compression-relaxation tests of granular columns that consist of dust particles or rigid glass beads. We found nonlinearly growing compression force both for dust-particles and glass-beads. Physics of the nonlinear growth in compression force is examined by comparing the models used in engineering fields. Furthermore, details of the compression-force behaviors depend on constituent particles. Dust-particles compression shows periodic undulation whose amplitude grows as compression proceeds. However, glass-beads compression exhibits quasi-periodic force drops due to the stick-slip rearrangement of grains network structure. To understand the physical nature of the observed dust-particles force undulation, fracturing (or deformation) of dust particles during the compression was also analyzed based on the image analysis. Concerning the relaxation process, the final equilibrium state is reached relatively faster in the dust particles than the glass beads because the former can easily be deformed and many of them are pulverized during the compression process. In this paper, we report on the analysis results of these compression-relaxation behaviors. Then, the physical meaning of the observed results is considered.

\section{Experiment}
\label{sec:experiment}
Simple compression-relaxation tests of a column consisting of hierarchical granular matter or glass beads were performed. To form hierarchical granular matter, dust particles were generated using a pan-type granulator (AS-ONE, PZ-01R). Tiny glass beads of diameter 5~{\textmu}m (Potters Ballotini, EMB-10, size range is 2--10~{\textmu}m) were poured on a pan and the pan was rotated. During the rotation, very small amount of water was sprayed on the rotated beads. Then, dust particles were formed by granulation of monomers (tiny glass beads). Monomers are presumably bound by humidity-induced micro capillary bridges and van der Waals force among monomer particles. The dust particles were sieved to collect particles with diameter $d$ in the range of $d=0.6$--$1.4$~mm. In this paper, we denote the typical particle diameter $d=1$~mm. Dust particles made by this procedure are stable once they are formed under the usual atmospheric conditions. The resulting particles from the above protocol approximately have grain-scale packing fraction (in the particles) $\phi_g \simeq 0.27$ (similar to \cite{Katsuragi:2018}) and their stiffness is sufficient to allow manipulation during the experiments. However, attrition of dust particles cannot be completely restrained.

\begin{figure}[ht!]
\begin{center}
\includegraphics[width=8.5cm]{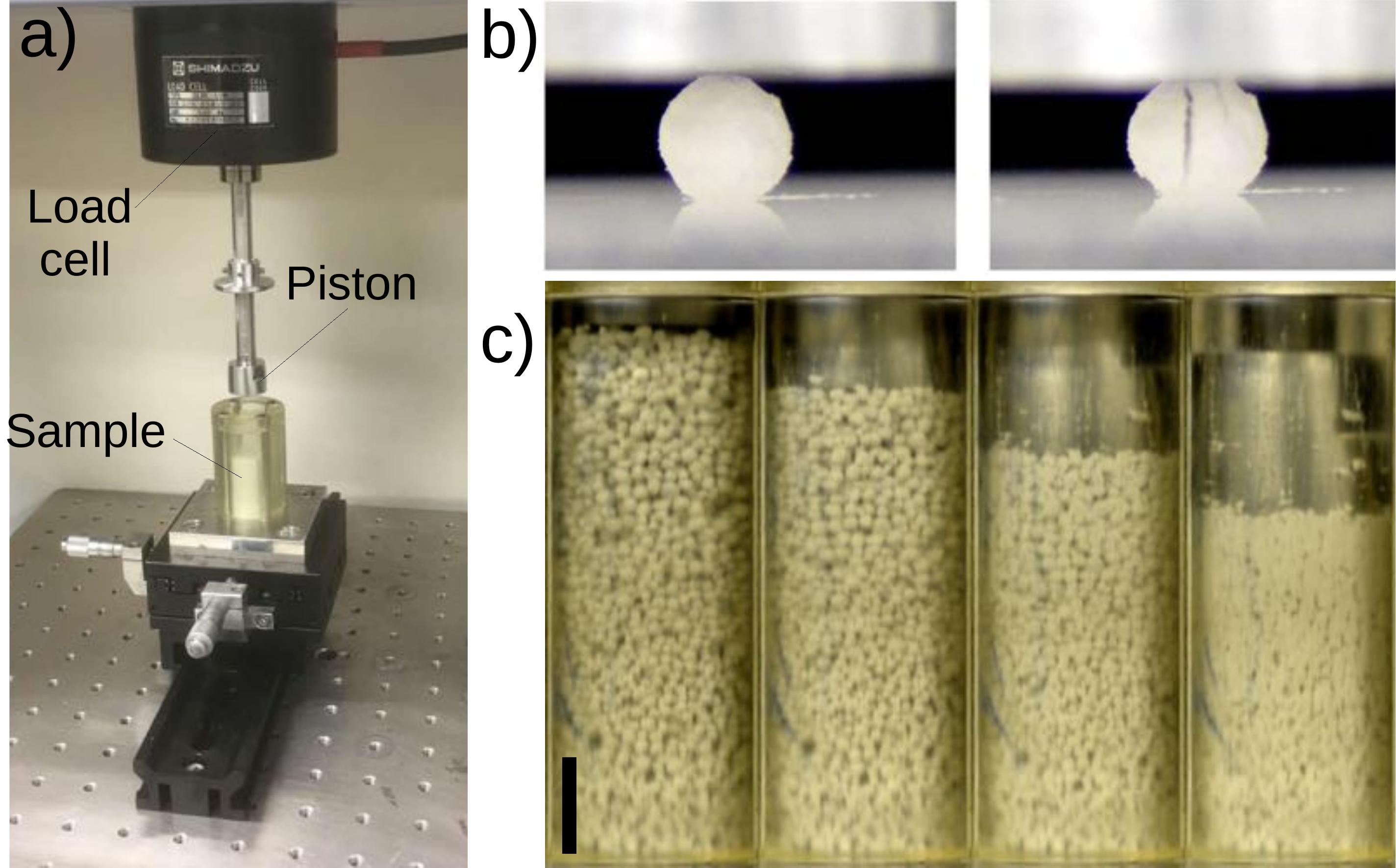}
\caption{a)~Experimental setup. b)~Compression of an individual dust particle showing the appearance of a fracture. c)~Snapshots showing the compression of a granular column composed of dust particles. 
}
\label{fig1}
\end{center}
\end{figure}

\subsection{Compression process} 
A cylindrical cell made of resin with inner diameter 20~mm and height 57~mm was mounted on a table of the universal testing machine (Shimadzu AG-X), see Fig. \ref{fig1}(a).  The cell was filled with dust particles. Specifically, fixed-mass dust particles  (8~g) are gently poured in the cylindrical cell~($\simeq 47$~mm height). Then, the dust-particles column was vertically compressed by an aluminium piston with a constant rate. The compression rate $v$ was precisely controlled by the testing machine and varied from 0.17 to 2000~\textmu m~s$^{-1}$ (over 4 orders of magnitude). Because a 50~N-capacity load-cell sensor was used in the experiment, the column was compressed until the compression force became $F_\mathrm{max}=20$~N. The vertical  compression stroke and force were acquired by 10 or 100 Samples~s$^{-1}$ sampling rate. Glass beads of diameter 0.7--0.9~mm (AS-ONE, BZ08) were also used for comparison. Mass of glass beads poured in the cell was fixed at 15~g to form a nearly fixed height column~($\simeq 33$~mm height). Basically, the identical experimental apparatus and protocol were used for the glass-beads experiments as well. 

Besides, we performed compression tests of a single dust particle. Pictures exemplifying the compression of an individual dust particle and a granular column of these grains (hierarchical granular matter) contained in the cylinder are shown in Figs. \ref{fig1}(b) and (c), respectively. Note in (b) how a vertical fracture visible to the eye appears due to the compression. This measurement allowed us to determine the failure stress (tensile strength) and the corresponding yield strain of dust particles, which are approximately 15~kPa and 0.05, respectively. Therefore, the apparent elastic modulus of dust particles is about $300$~kPa. These values are quite smaller than typical glass strength $\sim 7$~MPa and its elastic modulus $\sim 60$~GPa. See Appendix~\ref{sec:single} for details of single dust particle compression test. 

As can be seen in Fig.~\ref{fig1}(b), dust particles have approximately spherical shape with rough surfaces. Thus, the macroscopic packing state should be similar between dust particles and glass beads. However, the initial macroscopic packing fraction $\phi_m$ is roughly $0.6$ for glass beads and $0.8$ for dust particles; namely, initial bulk packing fraction of dust particles $\phi_\mathrm{bulk} = \phi_g \times \phi_m$ is about $0.22$. The relatively large $\phi_m$ value for dust particles comes from polydispersity induced by attrition during the handling. Although we carefully handled dust particles, it was difficult to prevent attrition. Therefore, detached small particles could fill macroscopic voids among large particles.  

The diameter of the cylindrical cell used in this experiment (20~mm) might not be large enough, particularly for $d=0.8$~mm glass beads. In general, wall friction, which might cause stick-slip motion and/or arch formation, could affect the compression force behavior. By increasing the diameter ratio between cell and particles, wall effect can effectively be reduced.  
To observe the effects of wall friction and hierarchical structure, a compression test of a granular column simply consisting of tiny glass beads (diameter 5~{\textmu}m) was also performed. 

As can be observed in Fig.~\ref{fig1}(c), the dust-particles column was significantly compressed and all the dust particles seem to be compressed more or less homogeneously. Thus, collective breakage of dust particles plays an essential role in compression force behavior. The collective particle breakage at the scale of the granular column allowed a visible compaction that can be recorded with a video camera [see snapshots in Fig. \ref{fig1}(c)], which was not possible for the glass beads. Digital Image Correlation analysis (DIC)~\cite{White:2003} was used to determine the displacement of grains between consecutive frames (velocity field) during the compression process.

\subsection{Relaxation process} 
Once the maximum compression force $F_\mathrm{max}$ was reached, the piston was stopped and kept in that position to study the force relaxation process. The relaxation of compression force was measured as a function of time. The experiments were performed using the dust particles and glass beads.  The room temperature, $T=25^\circ$C, and humidity,  $\sim 50\%$, were controlled and kept constant during each run.

\section{Results and analyses}
\begin{figure}[ht!]
\begin{center}
\includegraphics[width=7.5cm]{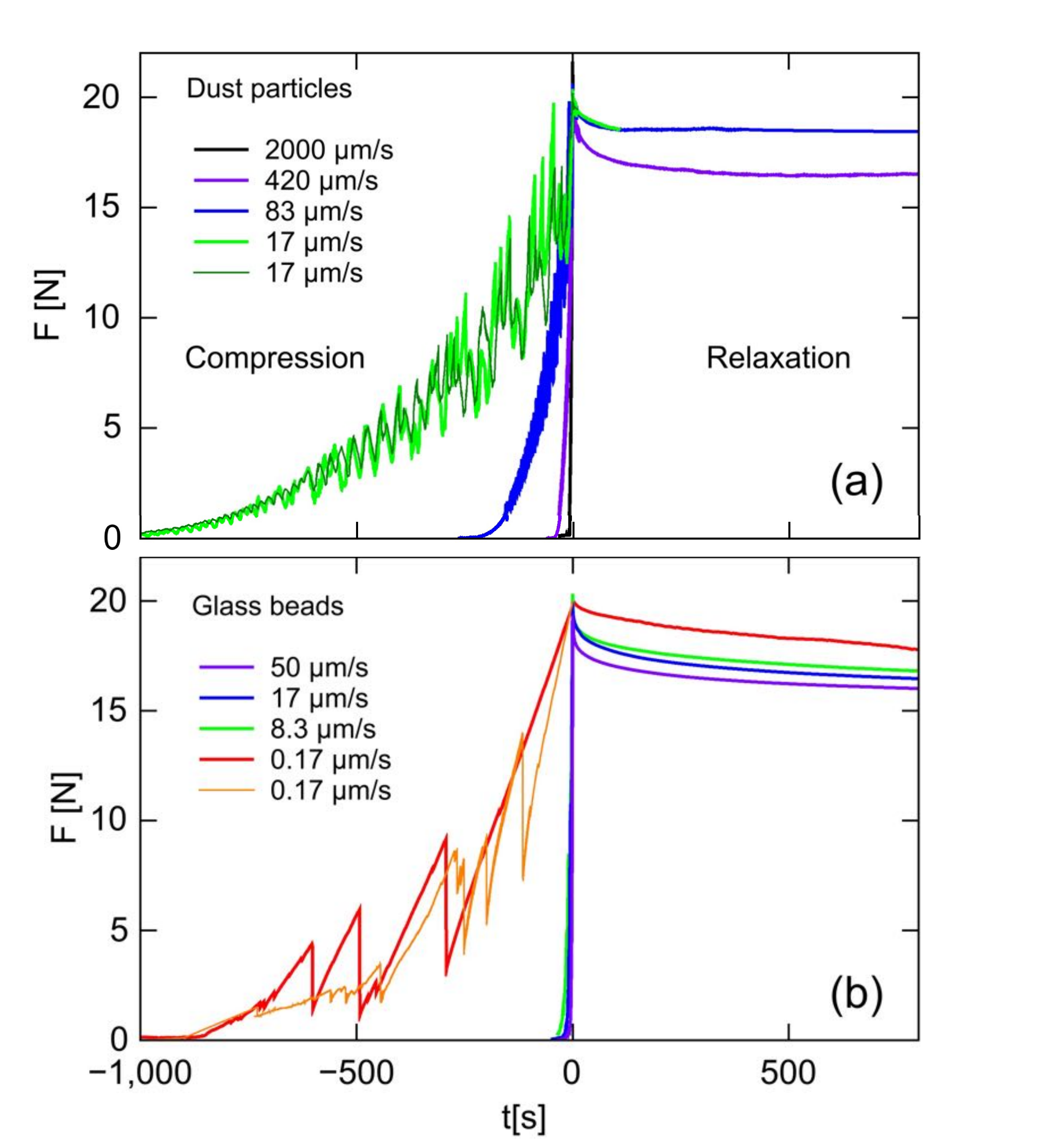}
    \caption{Compression force $F$ as a function of time $t$ measured at different compression rates followed by system relaxation for (a) dust particles and (b) glass beads. In these plots, $t=0$ corresponds to the moment at which the maximum compression load $F_\mathrm{max}=20$~N is attained by increasing the stroke $z$ from 0 to $z_\mathrm{max}$. After $t=0$, the stroke is fixed and the force is measured as a function of time $t$. 
}
\label{fig2}
\end{center}
\end{figure}

Figure~\ref{fig2} shows measurements of the force $F$ vs. time $t$ for granular columns composed of dust particles [Fig. \ref{fig2}(a)] and glass beads [Fig.~\ref{fig2}(b)] subjected to different compression rates. In the case of glass beads, only small values of compression rates were used because the stiffness of the particles produces a faster increase of the force load, in comparison with dust particles in which case we can use larger compression rates. In these plots, negative values of $t$ correspond to the compression phase and the positive ones to the relaxation phase. First, the compression process was performed by increasing the stroke $z$ from zero to a maximum value $z_\mathrm{max}$ ($t=0$~s) at which the maximum force $F_\mathrm{max}=20$~N was reached.  Then, the relaxation phase began and it was characterized by the continuous decrease of $F(t)$. For all the studied cases, the compression phase can be characterized by fluctuations of $F$, although the form and size of these fluctuations depend clearly on the nature of the granular particles. On the other hand, the relaxation phase is characterized by an initial exponential decay of $F(t)$ that depends on the compression rate, followed by a slow and smooth decrease of $F(t)$. Note that the relaxation to the equilibrium state was achieved faster by the dust particles than the glass beads. Because we would like to focus on the difference between dust particles and glass beads in force fluctuation and relaxation, we fixed most of the parameters. The principal parameter varied in this experiment was the compression rate $v$. To check the reproducibility of the experiment, slow compression tests ($17$~\textmu m~s$^{-1}$ for dust particles and $0.17$~\textmu m~s$^{-1}$ for glass beads) were repeated as shown in Fig.~\ref{fig2}.

Let us first focus on the compression of granular columns and then compare the relaxation process for both kinds of materials.

\subsection{Dust particles compression}
\begin{figure}[ht!]
\begin{center}
\resizebox{0.45\textwidth}{!}{\includegraphics{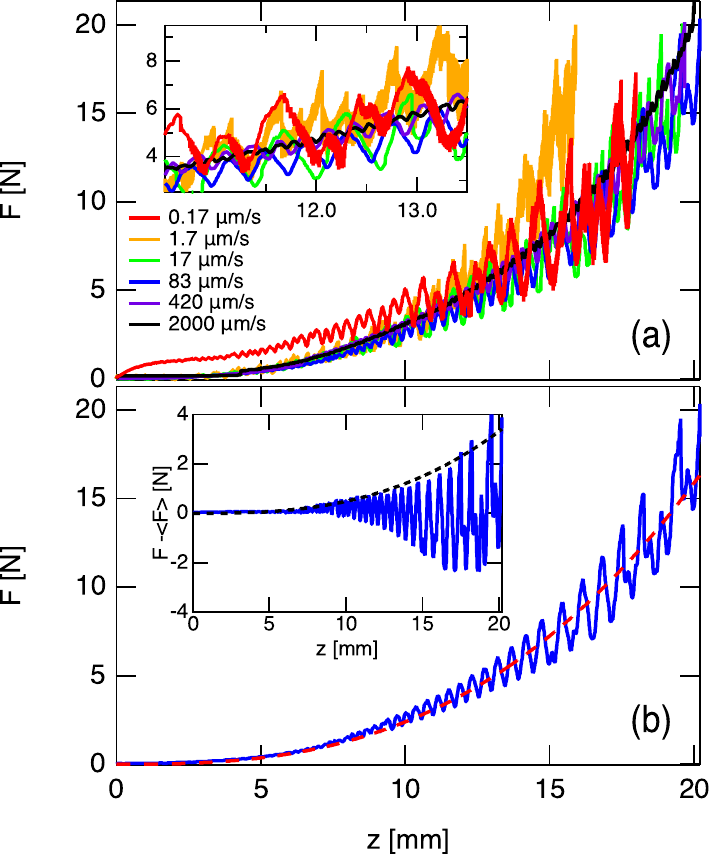}}
\end{center}
  \caption{Force $F$ vs. stroke $z$ for a dust-particles-column compression. Color in (a) indicates the compression rate $v$ as denoted in the legend. The inset of (a) displays the magnified view of periodic undulations. A typical example of $F(z)$ ($v=83$~{\textmu}m~s$^{-1}$)  is presented with its power-law fit (red dashed curve) in (b). The inset of (b) shows the development of the periodic undulation with a power-law envelope fit (black dotted curve). }
\label{fig:HG_raw}
\end{figure}

Compression data for dust particles are shown in Fig.~\ref{fig:HG_raw}. When we plot $F(t)$ as shown in Fig.~\ref{fig2}, it is difficult to see the details on data behavior in compression phase ($t<0$) because of the wide variation of compression rate $v$. Nevertheless, $v$-independent nature of $F$ can be realized when we plot $F(z)$. Therefore, the compression force $F$ is plotted as a function of compression stroke $z$ in Fig.~\ref{fig:HG_raw}. The level $z=0$ is manually determined by the contact of the piston and the column's top surface, and the downward (compressive) direction is set as positive direction of $z$. In Fig.~\ref{fig:HG_raw}(a), force curves with various compression rates are displayed. As can be seen, $F$ grows nonlinearly with increasing $z$. Although $F(t)$ curves for various $v$ values look quite different in Fig.~\ref{fig2}(a), the form of $F(z)$ is almost independent of $v$. This indicates that the compression proceeds in a quasi-static manner. 

Another prominent feature observed in the force curves is periodic undulation of $F(z)$. The inset of Fig.~\ref{fig:HG_raw}(a) shows the magnified view of the force curves. As seen in Fig.~\ref{fig:HG_raw}(a), amplitude of the undulation depends on $v$. The slower compression yields the larger amplitude. Wavelength of the undulation is in the order of mm and slightly depends on $v$.  

To analyze these characteristic features, we fit the mean growth of $F(z)$ to power-law function and extracted the undulation component. A typical example of this procedure is shown in Fig.~\ref{fig:HG_raw}(b), in which the force curve with $v=83$~{\textmu}m~s$^{-1}$ is shown. The red dashed curve indicates the power-law fitting of the mean growth of the compression force, $\langle F\rangle \propto z^{\alpha}$ with $\alpha=2.7$. The undulation component computed by $F-\langle F \rangle$ is shown in the inset of Fig.~\ref{fig:HG_raw}(b). The nonlinear growth of undulation amplitude can also be clearly confirmed. The black dotted curve in the inset of Fig.~\ref{fig:HG_raw}(b) indicates the power-law fitting of the envelope, $|F-\langle F \rangle|_\mathrm{env} \propto z^{\alpha_p}$ with $\alpha_p=2.7$. This type of power-law growth found in $F$ and $F-\langle F \rangle$ can be observed in all $F(z)$ curves of dust-particles compression. Thus, we fitted all the data with $\langle F \rangle=F_*(z/d)^{\alpha}$ and $|F-\langle F \rangle|_\mathrm{env} = F_{*p}(z/d)^{\alpha_p}$, where $F_*$, $F_{*p}$, $\alpha$, and $\alpha_p$ are fitting parameters.

\begin{figure}[ht!]
\begin{center}
\resizebox{0.45\textwidth}{!}{\includegraphics{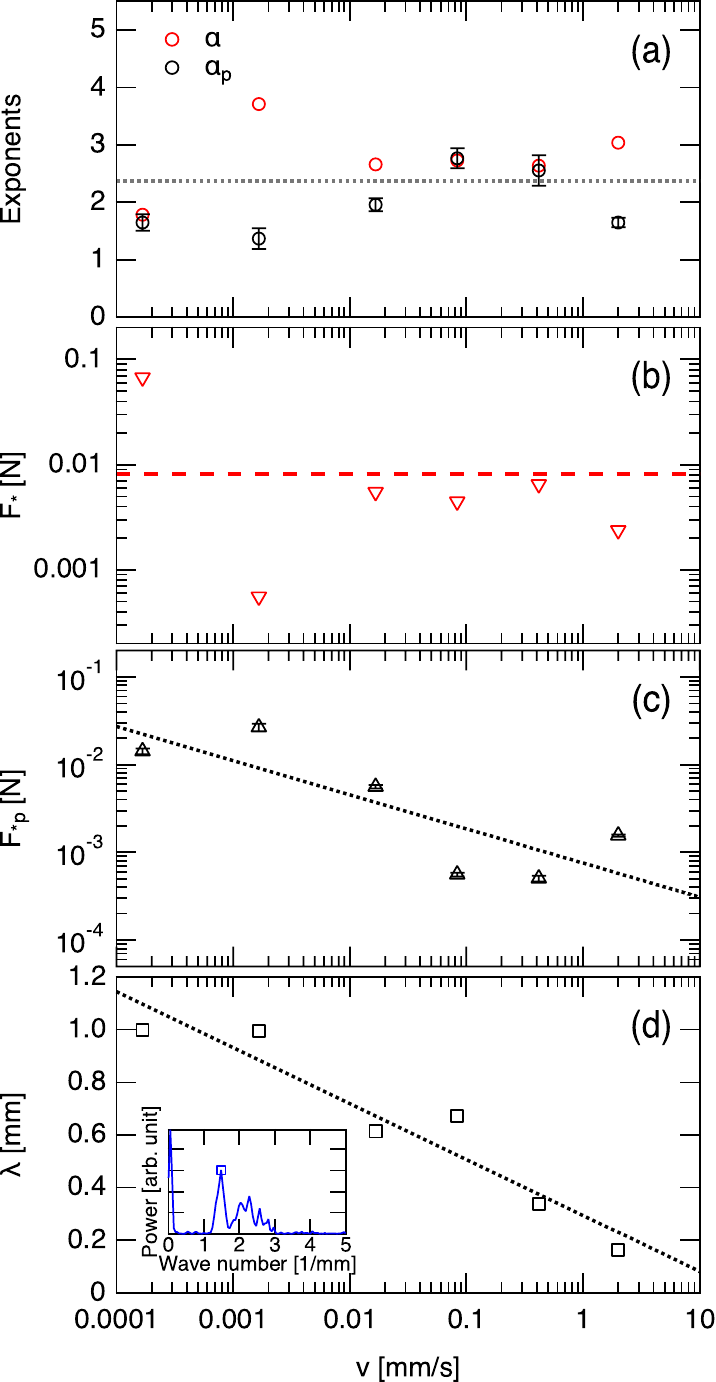}}
\end{center}
  \caption{Compression rate $v$ dependences of the parameters characterizing $F(z)$ curves in dust-particles compression. (a)~Power-law exponents for bulk compression and periodic undulation ($\alpha$ and $\alpha_p$) are almost independent of $v$; $\alpha \simeq \alpha_p = 2.4 \pm 0.7$. (b)~Force coefficient for bulk compression $F_*$ shows large fluctuation around $F_* \simeq 10^{-2}$~N. (c)~Force coefficient characterizing periodic undulation $F_{*p}$ shows power-law decay depending on $v$; $F_{*p} \propto v^{-\beta}$ with $\beta=0.4 \pm 0.1$. (d)~Dominant wavelength of periodic undulation $\lambda$ logarithmically decreases with increasing $v$; $\lambda = \lambda_* \ln(v_\mathrm{up}/v)$, where $\lambda_*=0.09 \pm 0.01$~mm and $v_\mathrm{up}=24 \pm 6$~mm~s$^{-1}$. The inset in (d) shows the power spectrum of the undulation shown in the inset of Fig.~\ref{fig:HG_raw}(b). Error bars indicate the fitting uncertainty. Error bars of red data points in (a) and (b) are omitted because the errors are smaller than the size of symbols.}
\label{fig:HG_params}
\end{figure}

The obtained fitting-parameter values as functions of $v$ are displayed in Fig.~\ref{fig:HG_params}(a-c). The power-law exponents $\alpha$ and $\alpha_p$ do not show any clear trend; they are rather almost constant (Fig.~\ref{fig:HG_params}(a)). Thus, we consider a unified value for the exponents $\alpha \simeq \alpha_p = 2.4 \pm 0.7$. For $F_{*}$, clear $v$ dependence cannot be confirmed again (Fig.~\ref{fig:HG_params}(b)). However, the data scattering particularly in small $v$ regime is significant. Thus, only the typical order of $F_*$ can be estimated as $F_* \simeq 10^{-2}$~N. The coefficient of force undulation $F_{*p}$ shown in Fig.~\ref{fig:HG_params}(c) exhibits a slightly negative correlation with $v$. The dotted line shown in Fig.~\ref{fig:HG_params}(c) is the power-law fitting, $F_{*p} \propto v^{-\beta}$ with $\beta=0.4 \pm 0.1$. This negative power-law correlation corresponds to the decrease tendency of undulation amplitude in the fast compression.  

Next, we computed the dominant wavelength of the periodic undulation. By identifying the peak wavenumber in the power spectrum of $F-\langle F \rangle$ (inset of Fig.~\ref{fig:HG_params}(d)), the dominant wavelength $\lambda$ can be easily computed. The obtained $\lambda$ values are plotted in Fig.~\ref{fig:HG_params}(d). The dotted line in Fig.~\ref{fig:HG_params}(d) indicates the logarithmic decay of $\lambda$, $\lambda = \lambda_* \ln(v_\mathrm{up}/v)$, where $\lambda_*=0.09 \pm 0.01$~mm and $v_\mathrm{up}=24 \pm 6$~mm~s$^{-1}$ were computed by the fitting. Since $\lambda$ becomes 0 at $v=v_\mathrm{up}$, $v_\mathrm{up}$ should correspond to the upper limit of the compression rate for the periodic undulation observation. However, we did not perform such a fast compression test due to the technical limitation.

By combining the above-mentioned analysis results, we finally arrive at the form of compression force,
\begin{equation}
\frac{F}{F_*} = \left( \frac{z}{d} \right)^{\alpha} \left[ 1 + \left( \frac{v}{v_1} \right)^{-\beta} \exp \left( \frac{2\pi i}{\lambda_* \ln (v_\mathrm{up}/v) }z \right) \right],
\label{eq:HG_F_form}
\end{equation}
where the value of $v_1$~($\simeq 10^{-2}$~mm~s$^{-1}$) can be estimated from the fitting results. This form includes all the characteristic features found in $F(z)$ curves of dust-particles-column compression.

\subsection{Glass beads compression}
\begin{figure}[ht!]
\begin{center}
\resizebox{0.45\textwidth}{!}{\includegraphics{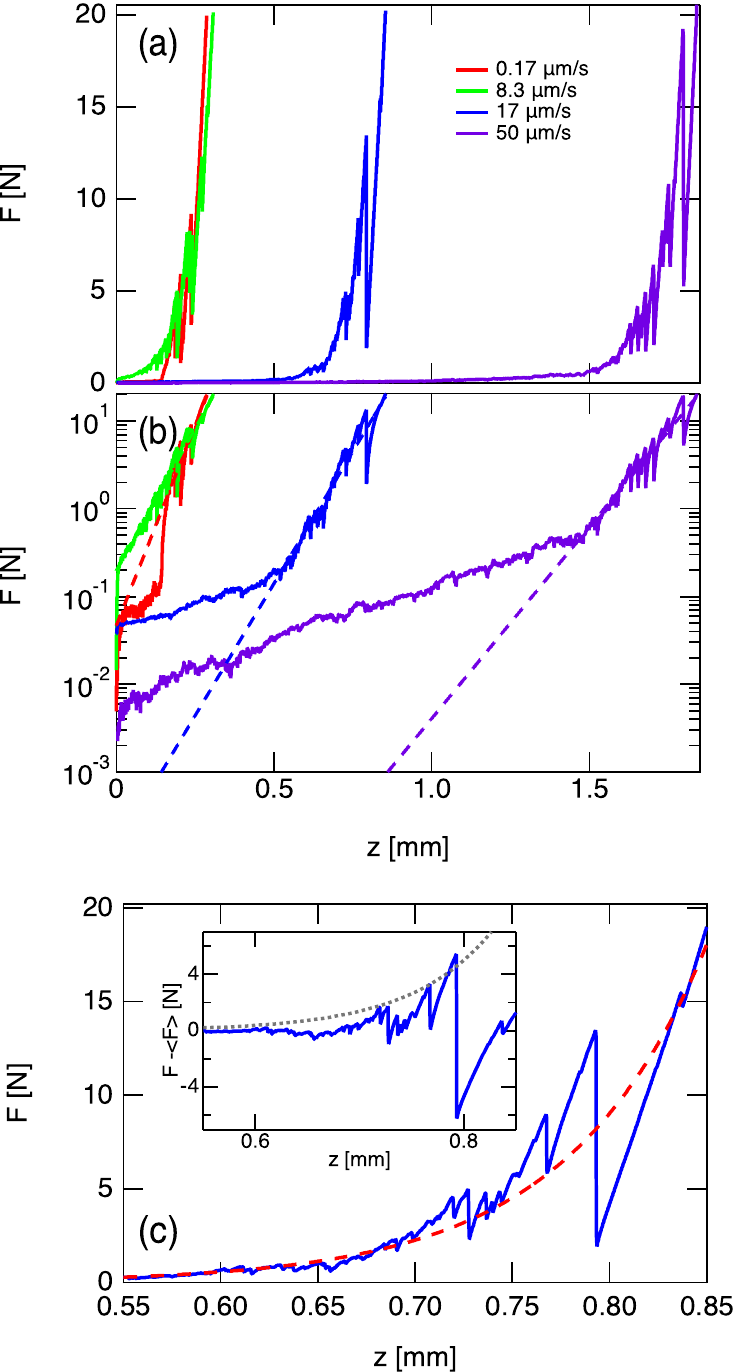}}
\end{center}
  \caption{Raw data set for glass-beads-column compression. $F(z)$ curves with various compression rates are plotted in panel (a). The panel (b) shows the log-linear plot of the same $F(z)$ data. $F(z)$ curve with $v=17$~{\textmu}m~s$^{-1}$ is presented in panel (c). The red dashed curve indicates the exponential fit. The inset in panel (c) shows the fluctuation part of the compression force with an exponential envelope fit (black dotted curve). }
\label{fig:GB_raw}
\end{figure}

To ensure the intrinsic nature of dust-particles compression, we performed similar compression tests with glass beads which are rigid enough in the current loading condition, $F\leq 20$~N. The results of glass-beads-column compression are shown in Fig.~\ref{fig:GB_raw}. $F(z)$ curves for glass beads with various compression rates are plotted in Fig.~\ref{fig:GB_raw}(a). The nonlinearity of $F(z)$ curves appears to be enhanced in the glass-beads compression. Figure~\ref{fig:GB_raw}(b) shows log-linear plot of the same data. One can confirm the clear exponential growth in the late stage of compression. Moreover, the range of $z$ to reach $F_\mathrm{max}=20$~N is $\sim 60$ times smaller than that of dust-particles compression. This rapid increase of $F$ results in its exponential growth rather than the power-law increase.

\begin{figure}[ht!]
\begin{center}
\resizebox{0.45\textwidth}{!}{\includegraphics{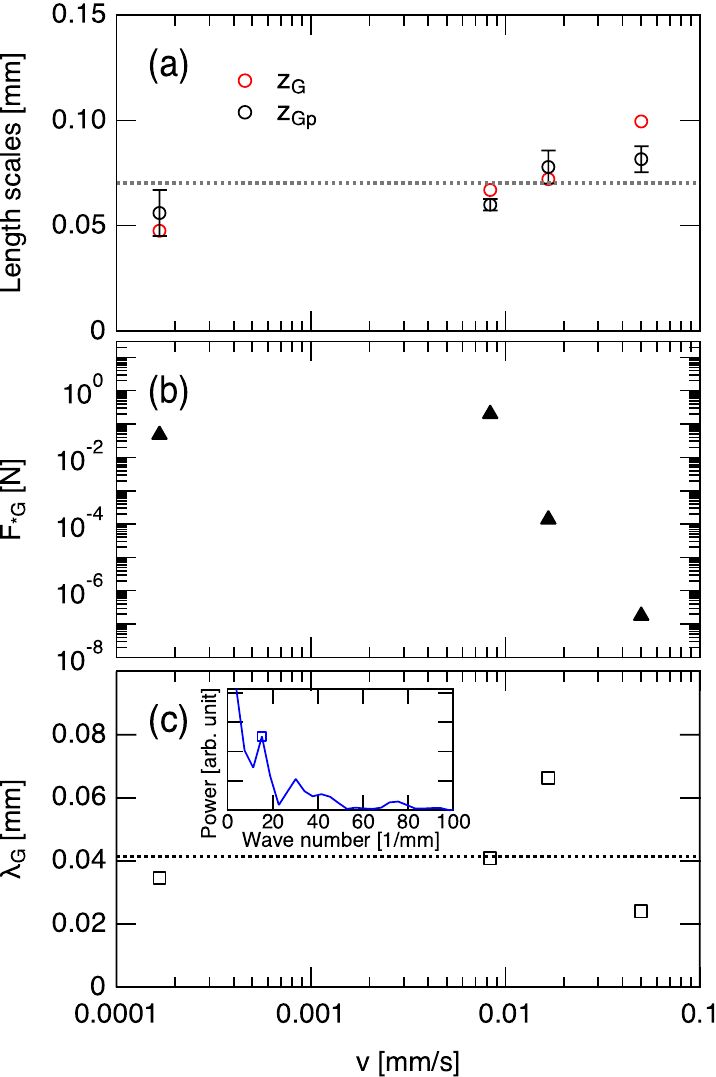}}
\end{center}
  \caption{The values of fitting parameters for glass-beads-column compression tests are presented as functions of the compression rate $v$. (a)~Characteristic length scales ($z_G$ and $z_{Gp}$) for exponential growths of bulk compression force and fluctuation amplitude are assumed to be roughly identical values independent of $v$; $z_{G} \simeq z_{Gp} = 0.07 \pm 0.01$~mm. (b)~The force coefficient value $F_{*G}$ is plotted as a function of $v$. Very large scattering of $F_{*G}$ (over 6 orders of magnitude) can be confirmed. (c)~The dominant wavelength $\lambda_G$ for the fluctuation part is also independent of $v$; $\lambda_G=0.04 \pm 0.02$~mm. The inset of (c) shows the power spectrum of the fluctuation shown in the inset of Fig.~\ref{fig:GB_raw}(d). Errors computed from the fitting uncertainty are shown if the scale of error is greater than the symbol's size.}
\label{fig:GB_params}
\end{figure}

A typical example of glass-beads compression with $v=17$~{\textmu}m~s$^{-1}$ is shown in Fig.~\ref{fig:GB_raw}(c). The red dashed curve represents the exponential fit. Although the notable fluctuation of $F$ can be observed in Fig.~\ref{fig:GB_raw}(c), its behavior is quite different from the periodic undulation seen in Fig.~\ref{fig:HG_raw}. In glass-beads compression, sudden force drops are repeated, whereas the periodic fluctuation can clearly be observed in dust-particles compression. The fluctuation component (force drops) in $F(z)$ was extracted also for glass-beads compression, as shown in the inset of Fig.~\ref{fig:GB_raw}(c). The growth of fluctuation amplitude is confirmed also in the glass-beads compression. The black dotted curve indicates the exponential growth of the fluctuation amplitude. By computing power spectrum of the fluctuation component (inset of Fig.~\ref{fig:GB_params}(c)), we can identify the dominant wavelength of the fluctuation $\lambda_G$ also in the glass-beads compression. 

The measured fitting parameters for glass-beads compression are shown in Fig.~\ref{fig:GB_params}. Here, we employ the exponential growth both for the mean force and its fluctuation as, $\langle F \rangle = F_{*G} \exp(z/z_G)$ and $|F-\langle F\rangle|_\mathrm{env} \propto \exp(z/z_{Gp})$, where $F_{*G}$, $z_{G}$, and $z_{Gp}$ are fitting parameters. In Fig.~\ref{fig:GB_params}(a), characteristic length scales, $z_{G}$ and $z_{Gp}$, are plotted as functions of $v$. Similar to the dust-particles compression, the growing manners of bulk compression force and its fluctuation amplitude coincide; $z_{G} \simeq z_{Gp} = 0.07 \pm 0.01$~mm. Whereas the slightly increasing trend can be confirmed in Fig.~\ref{fig:GB_params}(a), here we assume a constant behavior as a first-order approximation for the sake of simplicity. The compression rate dependence of $F_{*G}$ is presented in Fig.~\ref{fig:GB_params}(b). The measured $F_{*G}$ values show significant data scattering. As for the characteristic wavelength $\lambda_G$, the measured values do not show clear $v$ dependence (Fig.~\ref{fig:GB_params}(c)). The average wavelength is $\lambda_G=0.04 \pm 0.02$~mm. Note that the value of $\lambda_G$ is about one order of magnitude smaller than $\lambda$ (dust-particles case).

From the above-mentioned analyses, the compression force for glass-beads column can be written as,
\begin{equation}
\frac{F}{F_{*G}} = \exp\left( \frac{z}{z_{G}} \right) \left[ 1 + C_{G}\exp\left( \frac{2\pi i}{\lambda_G}z \right) \right]. 
\label{eq:GB_F_form}
\end{equation}
The parameter $F_{*G}$ is actually very sensitive to the initial condition. The value of $F_{*G}$ shown in Fig.~\ref{fig:GB_params}(b) varies over six orders of magnitude (see also the fitting lines shown in Fig.~\ref{fig:GB_raw}(b)). The large variation of $F_{*G}$ comes from the subtlety of initial response in the compressed glass-beads layer. By definition, compression begins when the piston reaches the particle on top. The subtle structure of grains configuration affects the behavior of early-stage compression of the glass-beads column since the glass beads cannot be deformed. Wide variation of the initial grains-network structure results in the large dispersion of the compression force in the early stage. To develop a stiff internal contact network, initial particle-level surface roughness must be flattened by compression. Figure~\ref{fig:GB_raw}(b) indicates that this preparation-related variation results in about one-particle-diameter scale fluctuation. Contrastively, the dust-particles column shows relatively good reproducibility because the particles can be easily deformed and yielded by the compression. Due to this deformability of particles, local deformation without network rearrangements is allowed. In very slow-compaction regime~($v\leq 0.002$~mm~s$^{-1}$), however, the data scattering of $F_*$ (Fig.~\ref{fig:HG_params}(b)) becomes significant even in dust-particles compression. Similarly, the value of $C_G$ in Eq.~\eqref{eq:GB_F_form} does not exhibit the universal behavior. Probably, these parameters are sensitive to the initial conditions (preparation dependence) of the column.  

Indeed, strong preparation (initial-condition) dependence is a universal feature of granular mechanics. For example, granular wall friction effect, so-called Janssen effect~\cite{Janssen:1895,Duran:2000}, strongly depends on preparation and initial condition~\cite{Katsuragi:2016}. However, this memory effect of the initial condition can be erased by adding small deformation to the granular system~\cite{Bertho:2003}. Therefore, the universal granular characteristic is observable after the small deformation. This tendency is more or less general in granular mechanics. Even in the simple simulation of granular compression test, careful preparation method~(e.g.~\cite{Taboada:2005}) is necessary to obtain reproducible results. Namely, we can consider that the large data scattering particularly observed in the force coefficient comes from the initial condition dependence. However, after slight compression, reproducible behaviors characterized by the universal parameters (length scales, time scales, power-law exponent, etc.) can be observed. In other words, global trend except the very initial stage is reproducible and universal. Namely, if we plot $F(z-z_\mathrm{max})$, the similar reproducible behaviors can be confirmed even for glass-beads compression.

\subsection{Image analysis of dust particles compression}
\label{sec:DIC}
\begin{figure*}[ht!]
\begin{center}
\includegraphics[width=18cm]{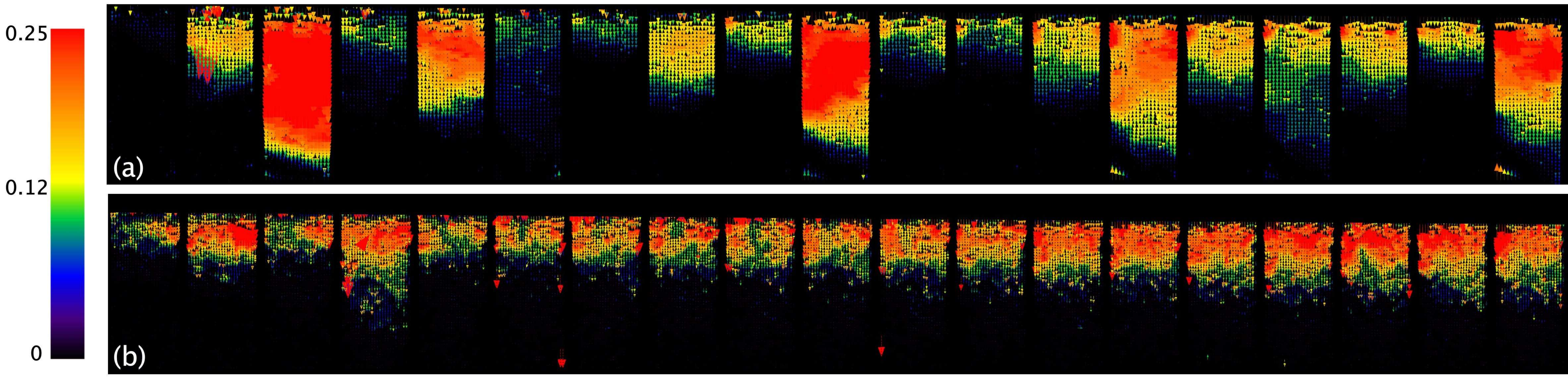}
\caption{Snapshots obtained from DIC analysis applied to consecutive frames of videos of dust particles compressed at (a) $v=17$~{\textmu}m~s$^{-1}$ and (b) $v=2000$~{\textmu}m~s$^{-1}$. The snapshots were taken at 0.073 frames~s$^{-1}$ and 5 frames~s$^{-1}$ in (a) and (b), respectively, and the scale color bar indicates the displacement in mm ranging from 0 to 0.25 mm of each vector (represented by a color arrow). The corresponding maximum velocity is 18 {\textmu}m~s$^{-1}$ and 1250 {\textmu}m~s$^{-1}$, respectively.  A three-step iterative analysis was applied considering interrogation window sizes of 512, 256 and 128 pixels in each case.
}
\label{DIC-analysis}
\end{center}
\end{figure*}

Next, we analyze the collective particle breakage that could relate to the periodic undulation observed in the dust-particles compression. Figure~\ref{DIC-analysis} shows snapshots obtained from DIC analysis of dust-particles compression at (a)~$v=17$~{\textmu}m~s$^{-1}$ and (b)~$v=2000$~{\textmu}m~s$^{-1}$. Only 18 frames are shown in each case, but they exemplify the behavior observed during all the compression processes. In these images, the upper black part corresponds to the piston position and each vector in color indicates the displacement of pixel patches between two consecutive frames. The color scale bar at the left of the figure indicates the corresponding displacement in mm. Note in Fig. \ref{DIC-analysis}(a) that at slow compression, there are frames practically colored in red, which indicates that all the granular column is rearranged as a whole in that precise moment. Then, the column displacement becomes negligible in the following frame and starts to increase again until a new general rearrangement occurs. We consider this general rearrangement event corresponds to the collective breakage of compressed dust particles. Note that this type of rearrangement is measurable only in dust-particles compression. 

If this collective breakage is the primary origin of the observed periodic undulation, wavelength shown in Fig.~\ref{fig:HG_params}(d) should agree with the stroke interval between consecutive general rearrangements (undulation wavelength $\lambda$). The number of frames between two consecutive general rearrangements obtained from analyzing the whole video is in average 3.9. Since the pictures were taken at 0.073 frames~s$^{-1}$, the rearrangement occurs approximately every 53 s, or equivalently, when the piston stroke increases about 0.9 mm. This value is close to the wavelength reported in Fig. \ref{fig:HG_params}(d) for the corresponding compression rate. Therefore, the periodic undulation of $F$ could relate to the general rearrangement of the granular column during the compression process. On the other hand, when the granular column was subjected to large compression rates, see Fig. \ref{DIC-analysis}(b), only a local rearrangement at the upper part of the granular column (close to the piston) was observed. This can be associated with the small amplitude observed for this compression rate in the experiments. Therefore,  at slow compression, the applied stress transmitted to the grains force them to enter into existing vacancies in the whole column and there is enough time for general rearrangement, but at fast compression, only the grains close to the piston have time to enter into the local vacancies and the rearrangement occurs only in that zone. The particle rearrangement can also be induced by particles breakage, in all the column for slow compression or localized below the piston at fast compression rates. This difference is reflected in the rate-dependent amplitude decrease found in dust-particles compression (Fig.~\ref{fig:HG_params}(c)).

While we can only observe the particle motion on the wall, we consider the similar behavior governs the whole column. Perhaps, wall-friction-induced slips or arch structure  might be developed. However, the relation between the global behavior shown in Fig.\ref{DIC-analysis} and the wall-induced slip or arch formation is quite unclear. Since we cannot observe the internal deformation, our understanding of collective particle breakage is limited. It should be noted, however, this breakage mode is quite different from conventional breakage observed in usual continuum materials.

\subsection{Relaxation of dust particles vs glass beads}
In the literature, different models have been proposed to describe the stress relaxation of a material previously subjected to a compression force. The generalized Maxwell model expresses the relaxation of stress $\sigma$ as a function of time  as: $\sigma(t) = \sigma_e + \Sigma_{j=1}^{n} C_j \exp(-t/\tau_j)$, where $\sigma_e$ is the equilibrium stress, $C_j$ are constant coefficients and $\tau_j$ are relaxation times.  For practical purposes, a three-term exponential equation has been found enough to describe the relaxation process in soft materials~\cite{Mohsenin:1978}. For a granular system subjected to oscillatory compaction that allows to achieve highly packed granular systems, the stress relaxation is found dependent on the strain rate. At slow strain rates, the relaxation is logarithmic in time, while at fast strain rates, the system follows a two-step relaxation~\cite{Brujic:2005} given by the expression: $\sigma(t)/\sigma_\mathrm{max} = A + B\exp(-t/\tau) + C\ln(t)$, where $\tau$, $A$, $B$, and $C$ are the relaxation time and fitting parameters; an instantaneous relaxation that decays exponentially followed by a logarithmic relaxation.

\begin{figure}[ht!]
\begin{center}
\resizebox{0.45\textwidth}{!}{\includegraphics{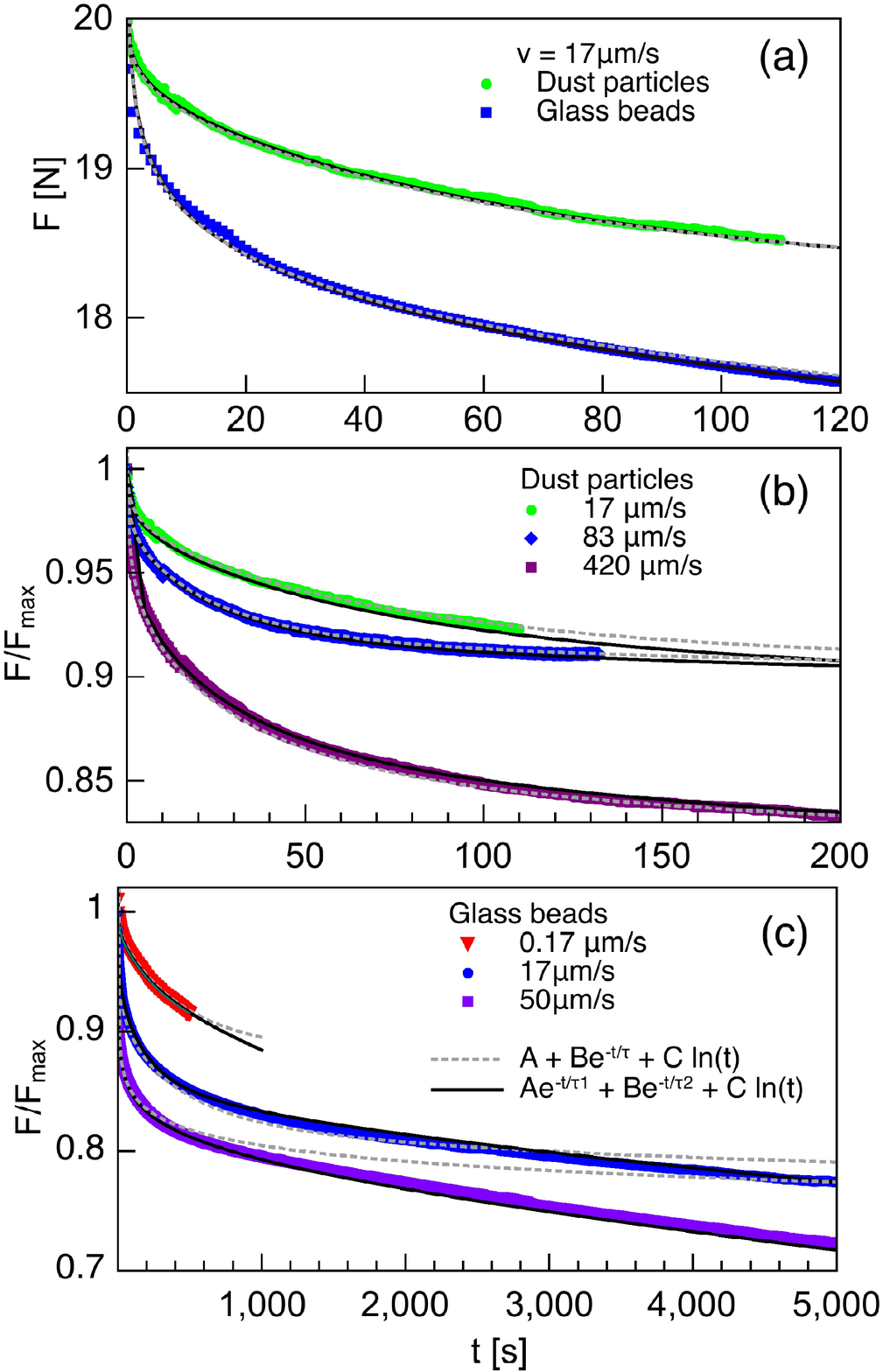}}
\end{center}
  \caption{(a) Direct comparison of $F/F_\mathrm{max}$~vs.~$t$ for dust particles and glass beads after being compressed until $F_\mathrm{max}=20$~N at $v=17$~{\textmu}m~s$^{-1}$. (b-c) $F/F_\mathrm{max}$ as a function of time for dust particles and glass beads, respectively. The black lines correspond to the data fit given by Eq.~\eqref{eq:Relaxation} with the parameters reported in Table~\ref{Table1}. Gray dotted curves show the fitting with $\tau_1=\infty$. }
\label{fig:relaxation}
\end{figure}

Let us analyze the relaxation process found in our experiments. Figures~\ref{fig2}(a) and (b) show $F$~vs.~$t$ for the relaxation of dust particles and glass beads, respectively. Whereas relaxations in very slow dust-particles-compression data ($v=1.7$ and 0.17~{\textmu}m~s$^{-1}$) could not be measured due the technical limitation, all the other available relaxation data were analyzed. We found that our data are not well described by the two step relaxation mentioned above particularly for glass-beads compression. Instead, it is necessary to consider an expression consisting of two exponential terms of considerably different time scale plus the logarithmic relaxation, in the form: 
\begin{equation}
\frac{F}{F_\mathrm{max}} =A\exp\left(-\frac{t}{\tau_1}\right) +B\exp\left(-\frac{t}{\tau_2}\right) + C\ln(t), 
\label{eq:Relaxation}
\end{equation}
where, $\tau_1$ and $\tau_2$ are characteristic relaxation times. The values of $\tau_1$, $\tau_2$ and the coefficients $A$, $B$, and $C$ obtained by fitting the experimental data shown in Fig.~\ref{fig:relaxation} (black curves) depend on the compression rate, and are reported in Table~\ref{Table1}. Equation~\eqref{eq:Relaxation} reduces to the one reported in \cite{Brujic:2005} (single exponential with constant and logarithmic terms) for considerably large values of $\tau_1$. This happens for dust particles, in which $\tau_1$ is three orders of magnitude larger than $\tau_2$. Therefore, the dust-particles relaxation curves can be fitted with the curves with $\tau_1=\infty$ (gray dotted curves in Fig.~\ref{fig:relaxation}(b).) However, single-exponential curves ($\tau_1=\infty$) cannot explain the glass-beads relaxation (gray dotted curves in Fig.~\ref{fig:relaxation}(c)).  From Table~\ref{Table1}, we can also notice that the main difference in the relaxation dynamics for dust particles and glass beads resides in the characteristic timescales, in particular in the fast relaxation characterized by $\tau_2$ for dust-particles compression. This means that the relaxation in a dust-particles column is faster than in a glass-beads column. Namely, we consider that while usual granular matter such as glass beads has a multi-stage relaxation, the longer timescale effect can be reduced due to the deformation and breakage in dust-particles compression. Because the force relaxation can be induced by particles rearrangement, multi-stage relaxation means multiple timescales in granular friction. This complexity of granular friction in a long timescale presumably relates to the complexity of friction itself~\cite{Kawamura:2012}. However, this multiplicity can be eliminated by particles deformation and breakage. Moreover, an instantaneous relaxation with $\tau_2\sim1~s$ is only found for glass beads when $v=50$~{\textmu}m~s$^{-1}$.  This fast relaxation observed at the largest compression rate would probably correspond to the two step relaxation for fast strain rates reported in Ref.~\cite{Brujic:2005}. Furthermore, longer time measurement is also required to precisely measure and discuss the longer relaxation time $\tau_1$. However, at the limit of $t\to\infty$, $F/F_\mathrm{max}$ diverges due to the $\ln(t)$ term in Eq.~\eqref{eq:Relaxation}. Thus, the relaxation model of Eq.~\eqref{eq:Relaxation} is just for the practical timescale phenomena. There could be more characteristic timescales $\tau_3$, $\tau_4$, $\dots$, if very long observation was possible. However, our measurements were restricted in practical timescale. A deeper analysis focused on the dependence of the parameters $\tau_1$, $\tau_2$, $A$, $B$ and $C$ on the compression rate, packing fraction, dust-particles stiffness, particle size and the maximum load are left for further research.

\begin{table}
\makegapedcells
\setlength\tabcolsep{5pt}
 \begin{tabularx}{\linewidth}{ c   c    c   c    c   c     }
    $v \, $({\textmu}m~s$^{-1}$)  &  A           &    B             & C                          & $\tau_1 \,$ (s)           & $\tau_2\,$  (s)       \\
  \Xhline{1pt} 
 {\bf Dust particles}:  \\ 
          17             &  0.942    &  0.043      & -0.0061     & $ \geq10^5 $          &      $72$     \\
          83            &  0.945    &  0.033      & -0.0072    & $ \geq10^5$          &   $26$        \\
          420          &  0.920    &  0.042      & -0.0150     &$ 45,000 $      &    $36$       \\
       \hline
{\bf Glass beads}:  \\
         0. 17             &  0.954    & 0.045        &-0.0048     & 29,000     &  290           \\
           8.3             &   0.945    & 0.025      &-0.0140      &54,000         & 270          \\
           17               &   0.950    & 0.043       &-0.0180     &70,000          & 300         \\
           50              &   0.936    & 0.050       &-0.0190      &80,000          & $0.5^*$        \\
           \hline
  \Xhline{1pt}   
\end{tabularx}
\caption{Values of the parameters used to fit the experimental data shown in Fig.~\ref{fig:relaxation}  using Eq.~\eqref{eq:Relaxation} for dust particles and glass beads in different compression rates. }
\label{Table1}
\end{table}

\section{Discussion}
In the compression phase, both dust-particles and glass-beads columns exhibit the nonlinear resistance force against the constant compression. However, the functional forms are different. Dust-particles and glass-beads compression obey power-law and exponential growths, respectively. This qualitative difference originates from deformability of the constituent particles. Dust particles can be easily deformed and broken while the glass beads particle cannot be significantly deformed. Thus, only the rearrangements of particle-network structure are permitted during the compression of glass-beads column. As a result, the rapid increase of bulk compression force with sudden force drops is observed in glass-beads compression. This implies that the fluctuation observed in glass-beads compression results from the stick-slip motion of contact network. Stick-slip behavior is a typical outcome of friction-induced slipping. Although we cannot distinguish wall-particle and particle-particle slips from the data, macroscopic sudden rearrangement of glass-beads network due to the frictional slip must be an origin of this behavior. In dust-particles compression, on the other hand, the local-compression-band formation and its propagation could occur as shown in Fig.~\ref{DIC-analysis}.  

Various models for the compression test of granular materials have been proposed. In soil mechanics, the compression index $C_c$ has been measured from the relation between void ratio $e=(1/\phi_\mathrm{bulk})-1$ and applied pressure $p=F/S$~\cite{Knapett:2012}, where $\phi_\mathrm{bulk}$ and $S$ are bulk packing fraction and area of compressing piston, respectively. We found that the exponential growth of $F(z)$ is consistent with this model, as shown in Appendix~\ref{sec:compaction_law}. In the field of powder engineering, however, Heckel model has been used to characterize the granular (void) compression~\cite{Denny:2002}. In the Heckel model, porosity $\varepsilon = 1-\phi_\mathrm{bulk}$ is related to its initial value $\varepsilon_0$ and applied pressure $p$ as $-\ln (\varepsilon) = \kappa_\mathrm{h} p - \ln(\varepsilon_0)$. Here, $\kappa_\mathrm{h}$ corresponds to a compressibility of voids. Although the Heckel model is partly consistent with our data, we consider the compression index model is better to explain the compression behavior, in terms of the applicable range. Details on the physical meaning of these models and the comparison with our experimental data are described in Appendix~\ref{sec:compaction_law}. 

However, these models cannot explain power-law growth observed in dust-particles compression. In this study, we empirically employed the power-law and exponential forms to simply show the coincidence of the growing manners between mean growth $\langle F \rangle$ and fluctuation growth $F-\langle F\rangle$. This correspondence is rather natural because both the mean growth of bulk compression force and fluctuation amplitude depends on $\phi_\mathrm{bulk}$ through the compression stress. In other words, the scale of plastic deformation and/or slips should relate to the scale of applied force. Furthermore, in some previous studies relating to dust compression, power-law relation between bulk packing fraction and compressive pressure has been assumed~\cite{Sirono:2004,Kataoka:2013,Omura:2018}. Although the situations (packing fraction regime and hierarchical structure) in these studies are different from ours, these results suggest there might be a universal power-law feature in easily-deformable dust compression. 

The substantial difference between glass beads and dust particles is deformability. In the glass-beads column, contact-network rearrangement is the only way to compress the layer. That is, the glass-beads column is almost in the jammed state even in its initial state. In such a jammed state, exponential growth of compression force is observed. If there is enough space to be compressed, power-law form can practically be applied in this state. Therefore, we expect the dust-particles layer should also obey the exponential form once it is significantly compressed (at least around jamming density). Namely, power-law form is an empirical one applicable to the early stage of very porous granular compression. In this experiment, however, we cannot compress the dust-particles column to this state due to the load cell capacity. Such a high compression test is required to conclude this expectation.

As for the periodic undulation, similar phenomenon (periodic fluctuation of granular drag force) was also found in plowing experiment~\cite{Gravish:2010}. In their experiment, the amplitude of the periodic fluctuation relates to the initial packing fraction of the granular layer. And the deformation of the free surface plays a crucial role to cause the periodic fluctuation. In this study, we confined particles in the cylindrical cell. Thus, any free-surface deformation is not allowed. However, there is still sufficient free space left for local deformation and compaction in dust-particles columns. As mentioned earlier, the initial bulk packing fraction of a dust-particles column is $\phi_\mathrm{bulk}\simeq 0.22$. Since $\phi_\mathrm{bulk}$ increases as the column is compressed, growth of undulation amplitude can be related to $\phi_\mathrm{bulk}$. However, contributions of $\phi_g$ and $\phi_m$ should be distinguished to discuss the analogy more precisely. To check the quantitative relation between the plowing and compression, much more careful observation in the compacted cell is needed. This problem is open to future. 

In order to understand the physical behavior of the compressed dust particles, the observed periodic undulation should be related to the collective breakage of dust particles. As demonstrated in Sec.~\ref{sec:DIC}, period (wavelength) of $F(z)$ undulation observed in dust-particles compression agrees with the interval of collective particle breakage. As shown in Figs.~\ref{DIC-analysis} and \ref{fig1}(c), the particle breakage propagates to the entire column. Similar compression wave propagation has been found in compressed crushable soft materials~\cite{Valdes:2011,Guillard:2015,Barraclough:2016,Golshan:2017}. Furthermore, more or less similar diffusive propagation of a compression band has been found in various conditions~\cite{Olsson:2001,Werther:2006,Holcomb:2007,Sheikh:2008,Tembe:2008}. Therefore, we consider that the particle breakage and its propagation are the main sources of the periodic undulation found in compressed dust-particles. Although its compression-rate dependence was analyzed and formulated (Eq.~\eqref{eq:HG_F_form}), fully convincing physical rationale of the obtained form has not yet been obtained. In this sense, our understanding of periodic undulation is still insufficient. 
According to \cite{Guillard:2015}, competition of the timescales is the principal source of oscillation found in the compression of crushable materials. However, this idea cannot be applied to the current case. In our experiment, the compression rate $v$ significantly varies over four orders of magnitude and the relaxation time $\tau_2$ is almost independent of $v$ (Table~\ref{Table1}). Nevertheless, we observed the undulation with almost a constant $\lambda$. Timescale-independent modeling is necessary for the reasonable explanation. We consider the physical mechanism causing periodic undulation in dust-particle compression is partly similar to but somewhat different from the phenomena reported in \cite{Valdes:2011,Guillard:2015,Barraclough:2016,Golshan:2017}. Systematic experiments under various conditions have to be performed to reveal the physical nature of this phenomenon.  
In addition, numerical simulations should also be performed to better understand the situation. Thus far, usual granular simulations have focussed on the discrete element methodology to analyze the structure and force chains of rigid-particle system~(e.g.~\cite{Taboada:2005}). Recently, a numerical method that can model the hierarchical granular matter has been developed~\cite{Jha:2020}. Systematic numerical simulation using such a new methodology is also helpful to reveal this novel phenomenon.

The remarkable feature found in the periodic undulation of dust-particles compression is the logarithmic dependence of wavelength $\lambda$ on the compression rate $v$. Perhaps, this logarithmic rate dependence may originate from the rate-state-dependent friction law~\cite{Marone:1998,Kawamura:2012}. Indeed, temporally logarithmic relaxation is observed as modeled in Eq.~\eqref{eq:Relaxation}. Moreover, the logarithmic relaxation is crucial particularly for dust particles, as discussed later. However, the observed undulation cannot be controlled by the timescales introduced by compression rate and force relaxation. Almost constant (at least in the same order) wavelength suggests spatial or geometrical conditions govern the undulation. To properly understand the undulation mechanism, the internal structural development of the compressed dust-particle column should be directly observed and analyzed. Unfortunately, such observation is not feasible at this time. Details on these logarithmic properties are crucial question opened to future.

Actually, we also performed single dust-particle compression tests (see Appendix~\ref{sec:single}) and a compression test of tiny monomers layer without granulation (see Appendix~\ref{sec:dust_layer}). However, both tests did not show any periodic undulation. Recently, compression tests of aggregates formed by mixture of snow, water, and tephra particles were also performed to discuss the mechanical properties of complex aggregates~\cite{Niiya:2020}. In the experiment, any periodic undulation was not found. Only the complex aggregation structure is not sufficient to produce periodicity. It can only be said that the hierarchical granular structure is necessary to induce periodic undulation. To discuss the origin of periodic undulation and its precise rate dependence, further compression tests and their statistical analysis are necessary. 

Regarding the relaxation process, the slower relaxation at long timescales (larger values of $\tau_1$) in the case of dust particles can be explained considering that these conglomerates are deformed and many of them pulverized during the compression process. Then, the material is closely packed and gets stuck due to the cohesion effect and the further rearrangement of the contact network takes longer time to occur. This is similar to what happens with the jammed state obtained when the system of glass beads is vibrated in Ref. \cite{Brujic:2005}. For that reason, the model given in Ref. \cite{Brujic:2005} is also able to predict the relaxation of dust particles. On the other hand, the glass beads remain practically intact after compression, which is reflected in a larger decrease of $F$ and shorter values of $\tau_1$. On the other hand, the short values of $\tau_2 < 10^2~s$ for dust particles imply that the logarithmic relaxation starts to be important first for this kind of material. This is contrastive with the glass beads where the logarithmic relaxation is only observed for the largest compression rate, which indicates that the glass beads continue rearranging even at very longer time. The role of friction with the container and between particles could also be playing an important role in this particle rearrangement.

In this experiment, we fixed most of the parameters: particle size ($d\simeq 1$~mm), cylindrical cell diameter (20~mm), and constituent particle material (glass). These factors could affect the behavior of compression-force fluctuation and its relaxation. In this study, by fixing these parameters, we have concentrated on the effect of hierarchical structure on the fluctuation of compression force and its relaxation. To completely understand the physical nature of granular compression, other parameters should also be varied and the obtained forms (Eqs.~\eqref{eq:HG_F_form},\eqref{eq:GB_F_form},\eqref{eq:Relaxation}) should be improved. Some parameter dependences will be presented elsewhere in near future~\cite{Pacheco:2021}.

\section{Conclusion}
We experimentally compared the compressive reaction and force relaxation of dust-particles column and glass-beads column. Both granular columns exhibit the nonlinear increase of compression force. In dust-particles compression, periodic undulation can be observed in addition to the nonlinear growth. In glass-beads compression, however, sudden force drops rather than undulation were observed during the compression. Parameters characterizing the compression-force behaviors were systematically analyzed using the experimental data. As a result, we found that most of the characteristic quantities are roughly independent of compression rate while the compression rate was varied over four orders of magnitude. The amplitude and wavelength of the periodic undulation found in dust-particles compression depend on the compression rate. Regarding the relaxation process, multiple time-scale relaxation followed by logarithmic relaxation was observed. We found these nontrivial phenomena for the first time by carefully comparing compressive reaction of dust particles and glass beads. 

By combining all the analyzed results, we introduced empirical models and discussed the corresponding physical meaning. Specifically, compression-force behaviors were modeled by Eqs.~\eqref{eq:HG_F_form} and \eqref{eq:GB_F_form} for dust-particles and glass-beads compression, respectively. In addition, force relaxation in the compressed granular layers were measured and modeled by Eq.~\eqref{eq:Relaxation}. Physical meanings of the obtained result were discussed based on various previous studies in the fields of physics, planetary science, soil mechanics, and powder engineering. Particularly, we found that the exponential force growth observed in glass-beads compression is consistent with the compression index model used in soil mechanics. Although the physical understanding of these behaviors was not completely established, reasonable understanding was obtained by considering the physical difference between dust particles and glass beads.

\begin{acknowledgments}
We thank JSPS KAKENHI Grant No.~18H03679 and Nagoya University for financial support. 
\end{acknowledgments}

\appendix
\section{Single particle compression test and stress comparison}
\label{sec:single}
\subsection{Single particle compression test}
\begin{figure}[ht!]
\begin{center}
\resizebox{0.45\textwidth}{!}{\includegraphics{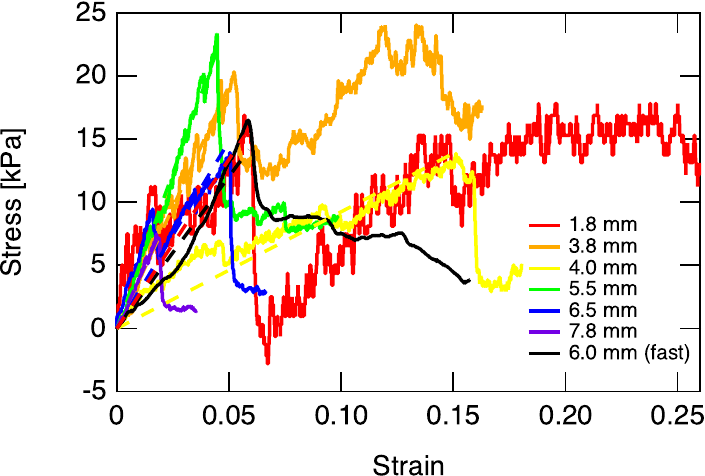}}
\end{center}
  \caption{Stress-strain relations for single dust particle compression tests. Particle size (diameter) was measured with a vernier caliper. Color indicates the particle size as labeled in the legend. Compression rate is basically fixed to $v=1.7$~{\textmu}m~s$^{-1}$ except for $6.0$~mm size particle case ($v=17$~{\textmu}m~s$^{-1}$).}
\label{fig:SingleCompression}
\end{figure}

To quantitatively assess the mechanical strength of dust particles, we performed single-particle compression tests. Because dust particles are very weak, it is difficult to pickup and handle a single small particle. Due to this technical limitation in handling dust particles, we compressed relatively large dust particles produced by the protocol mentioned in Sec.~\ref{sec:experiment}. Dust particles of diameter $d_p=1.8$--$7.8$~mm (measured by a vernier caliper) were compressed directly by the piston (without sidewall, see Fig.~\ref{fig1}(b)). The compression rate was set $v=1.7$~{\textmu}m~s$^{-1}$ except one trial ($v=17$~{\textmu}m~s$^{-1}$ only for 6.0~mm diameter particle). Compression force and stroke must be appropriately normalized to obtain the mechanical property. Therefore, here the relation between stress and strain is considered. To evaluate stress, the compression force $F$ was divided by the cross-sectional area of particle $\pi d_p^2/4$. For the definition of strain, stroke $z$ is simply divided by $d_p$. The measured stress-strain relations are shown in Fig.~\ref{fig:SingleCompression}. As expected, any systematic size dependence cannot be observed. As can also be seen, the linear stress-strain relation continues upto about 5\% strain. Then, the fracturing could occur around 15~kPa stress. It should be noted that the dust particles are basically plastic than elastic. Thus, the plastic deformation is induced even in the initial linear compression regime. Anyway, we can roughly capture the mechanical properties of dust particles from this experiment.

\subsection{Stress comparison with periodic undulation}
\begin{figure}[ht!]
\begin{center}
\resizebox{0.45\textwidth}{!}{\includegraphics{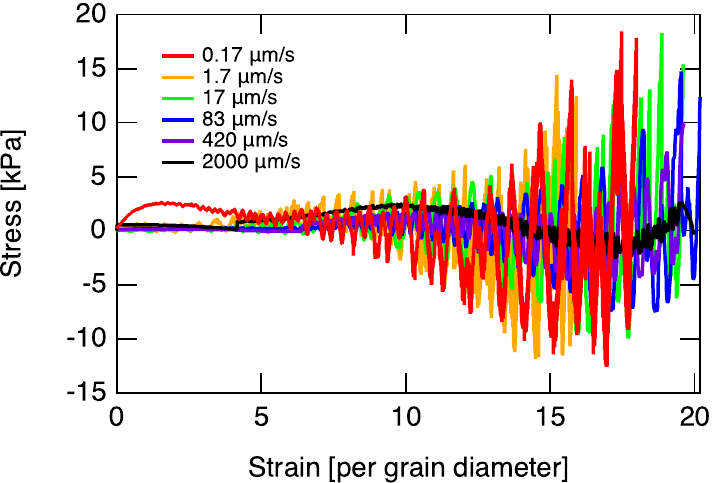}}
\end{center}
\caption{Stress-strain relations for periodic undulations in dust-particles-column compression. The color code used is identical to Fig.~\ref{fig:HG_raw}(a). Definition of strain is same as the single particle compression.}
\label{fig:HG_stress}
\end{figure}

In Fig.~\ref{fig:HG_stress}, stress of the periodic undulation part for dust-particles compression, $(F-\langle F \rangle)/S$ computed from the data shown in Fig.~\ref{fig:HG_raw}, is shown as a function of strain per grain diameter, $z/d$ (with $d=1$~mm). The undulation amplitude reaches about the mechanical strength of dust particles around $10d$ compression. We are not sure whether the upper limit of the amplitude of periodic undulation is restricted by the strength of particles, or not. Much more compression is necessary to reveal the behaviors of highly compressed dust-particles column. However, the early stage of compression cannot be resolved when we focus on the highly compressed stage by using a large capacity load cell. In this study, we have focussed on the early stage (lightly loaded state) of compression of a granular column. The heavy compression test is a possible future work.

\section{Compression laws}
\label{sec:compaction_law}
\subsection{Bulk modulus}
In order to systematically compare various types of granular-compression models, we consider the effective bulk modulus $K$ defined as
\begin{equation}
K= -V\frac{dp}{dV},  
\label{eq:bulk_modulus_def}
\end{equation}
where $p=F/S$ is pressure and $V=S(z_0-z)$ is volume of the compressed column, i.e., $z_0$ is the initial height of the column. The right-hand side of Eq.~\eqref{eq:bulk_modulus_def} can be written as $[(z_0-z)/S](dF/dz)$. For dust-particles compression, we assume power-law growth of $F(z)$ which can be expressed as $dF/dz=\alpha F/z$. Thus, $K$ for power-law form (dust-particles compression) becomes
\begin{equation}
  K_\mathrm{power}= \alpha p\left(\frac{z_0 -z}{z} \right).
  \label{eq:powerlaw_K}
\end{equation}
For exponential growth of glass-beads compression, the corresponding differential equation is written as $dF/dz=F/z_G$. Thus, $K$ should be 
\begin{equation}
  K_\mathrm{exp}= p\left(\frac{z_0-z}{z_G}\right).
  \label{eq:exponential_K}
\end{equation}
Here we consider the compression with ideal gas to develop the physical image of bulk modulus formulation. Adiabatic ideal gas satisfying $pV^{\gamma}=\mathrm{const.}$ yields $K_\mathrm{gas}=\gamma p$, where specific heat ratio $\gamma$ is constant. By comparing this relation with Eqs.~\eqref{eq:powerlaw_K} and \eqref{eq:exponential_K}, it can be said that the granular bulk modulus is sensitive to its column height (volume) than gas.

\subsection{Compression index}
\begin{figure}[ht!]
\begin{center}
\resizebox{0.45\textwidth}{!}{\includegraphics{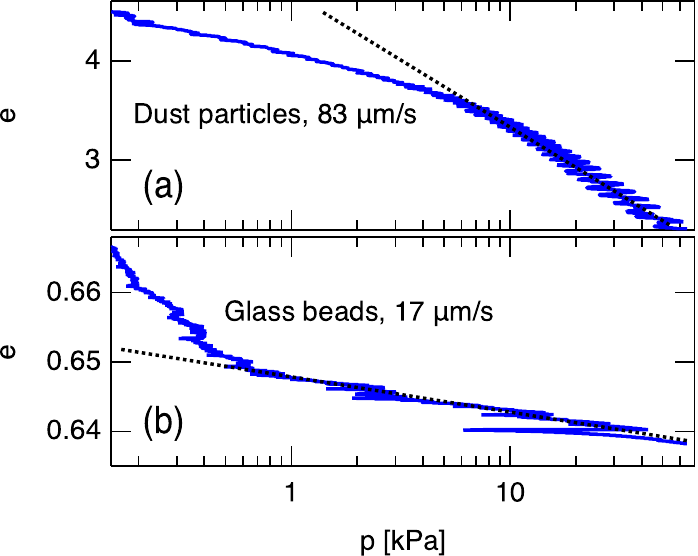}}
\end{center}
\caption{Plot for evaluating the compression index model. Data from (a) dust-particles compression and (b) glass-beads compression are presented. The data are identical to those shown in Fig.~\ref{fig:HG_raw}(b) and Fig.~\ref{fig:GB_raw}(b). Dotted lines are displayed as guides to the eye. The steep angle in small $p$ of (b) comes from the subtlety in the initial stage of glass-beads compression (Fig.~\ref{fig:GB_raw}).}
\label{fig:e_logp}
\end{figure}

Next, we introduce the compression index which has been used to characterize soil-compression characteristics~\cite{Knapett:2012}. Void ratio $e$ is defined by the ratio between void volume $V_\mathrm{void}$ and solid volume $V_\mathrm{solid}$, $e=V_\mathrm{void}/V_\mathrm{solid} = (1/\phi_\mathrm{bulk})-1$. Then, the compression index $C_c$ is defined by the relation between $e$ and $p$ as 
\begin{equation}
\frac{de}{d (\ln p)} = - C_c.  
\label{eq:compression_index}
\end{equation}
Specifically, the slope in $e$~vs.~$\ln p$ corresponds to $C_c$. The corresponding plot is shown in Fig.~\ref{fig:e_logp}. One can confirm the wider linearity in Fig.~\ref{fig:e_logp}(b) than (a). This is in fact because Eq.~\eqref{eq:compression_index} is consistent with Eq.~\eqref{eq:exponential_K} as explained below. By considering the above-mentioned relations and $\phi_\mathrm{bulk}=M_\mathrm{tot}/{\rho_\mathrm{true}V}$, where the total mass $M_\mathrm{tot}$ and true density $\rho_\mathrm{true}$ are constant, bulk modulus becomes
\begin{equation}
K_\mathrm{C} = \frac{\rho_\mathrm{true}S}{C_c M_\mathrm{tot}}p(z_0-z).
\label{eq:Cc_K}
\end{equation}
This relation provides a physical meaning of the length scale $z_\mathrm{G}$. In other words, the exponential form of $F(z)$ means that the compression index is a constant. For dust-particles compression, linear region is restricted as shown in Fig.~\ref{fig:e_logp}(a). This is the reason why the power law fits better than the exponential form for dust particles. However, the linearity in the late stage ($p \geq  10$~kPa) is reasonable in Fig.~\ref{fig:e_logp}(a). That is, the compression index can properly express the late-stage physical behaviors in both glass-beads and dust-particles compression. The crossover point ($\sim 10$~kPa) roughly coincides with the strength of a single particle.

\subsection{Heckel model}
\begin{figure}[ht!]
\begin{center}
\resizebox{0.45\textwidth}{!}{\includegraphics{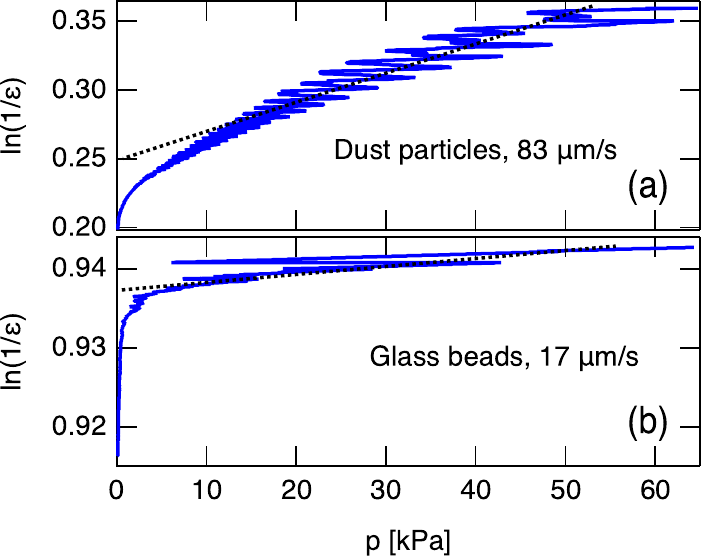}}
\end{center}
\caption{Heckel model plot of (a) dust-particles compression and (b) glass-beads compression. The data are identical to those shown in Fig.~\ref{fig:HG_raw}(b) and Fig.~\ref{fig:GB_raw}(b). Dotted lines are shown as guides to the eye.}

\label{fig:Heckel}
\end{figure}

Another popular model characterizing granular compression is Heckel model~\cite{Denny:2002}. The model considers the compressibility of voids. The model is written as,
\begin{equation}
-\frac{1}{\varepsilon} \frac{d\varepsilon}{dp} = \kappa_\mathrm{h},
\label{eq:Heckel_def}
\end{equation}
where $\varepsilon$ and $\kappa_\mathrm{h}$ are porosity ($\varepsilon=1-\phi_\mathrm{bulk}$) and compressibility of voids, respectively. By solving Eq.~\eqref{eq:Heckel_def},  the relation $-\ln \varepsilon = \kappa_\mathrm{h} p - \ln(\varepsilon_0)$ is obtained, where $\varepsilon_0$ is the initial value of $\varepsilon$. To evaluate the validity of this model, linearity of $\ln(1/\varepsilon)$~vs.~$p$ is usually inspected. The corresponding plot is shown in Fig.~\ref{fig:Heckel}. Although the data obey the Heckel model particularly in large $p$ regime, the linearity is not very good. Note that the linear abscissa in Fig.~\ref{fig:Heckel} and logarithmic abscissa in Fig.~\ref{fig:e_logp}. Both the linearity and linear ranges in Fig.~\ref{fig:Heckel} are inferior to those in Fig.~\ref{fig:e_logp}. Thus, we consider that our data can be better characterized by the compression-index model than by Heckel model. Since the compressibility is simply a reciprocal number of bulk modulus, we can readily obtain the corresponding bulk modulus form as 
\begin{equation}
  K_\mathrm{h} = \frac{1}{\kappa_\mathrm{h} e}.
\end{equation}
This form is different from Eqs.~\eqref{eq:powerlaw_K} and \eqref{eq:exponential_K}.

\begin{figure}[ht!]
\begin{center}
\resizebox{0.45\textwidth}{!}{\includegraphics{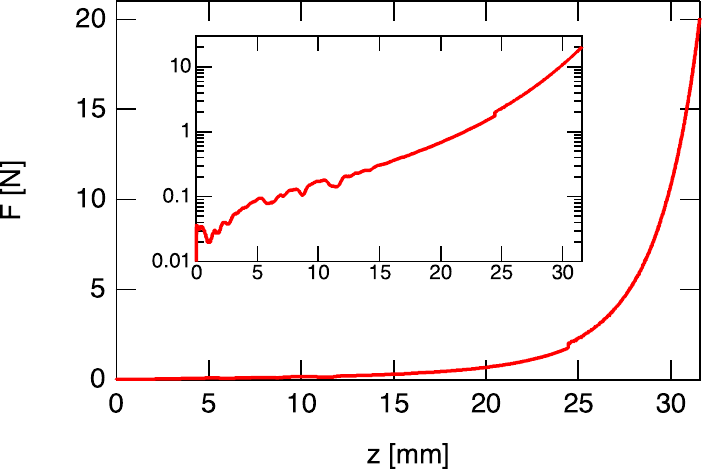}}
\end{center}
\caption{Force curve of the compression test of monomers. Tiny ($5$~{\textmu}m) glass beads were used as monomer particles ($\phi_\mathrm{bulk} \simeq 0.3$). Monomers are simply poured into the cylindrical cell without granulation process. Neither periodic undulation nor sudden force drops are observed during the compression. The inset shows the log-linear plot of the same data.}
\label{fig:Dust_raw}
\end{figure}

\section{Compression of a dust layer}
\label{sec:dust_layer}
The simple compression test of a granular layer consisting of tiny monomers ($5$~{\textmu}m glass beads) was also performed to check the importance of a hierarchical structure for causing periodic undulation. Furthermore, the effect of wall friction could be relatively reduced because the diameter ratio between the cell and particles becomes very large (4000) in this setup. In this test, monomers were directly poured in the cylindrical cell (without granulation process), and the column was compressed with $v=0.42$~mm~s$^{-1}$. Note that the experimental setup is identical to the dust-particles compression case. Bulk packing fraction of this column was $\phi_\mathrm{bulk} \simeq 0.3$. This value is close to dust-particles packing fraction $\phi_g=0.27$. The obtained result ($F(z)$ curve) appears in Fig.~\ref{fig:Dust_raw}. We can observe the nonlinear growth of $F(z)$. The inset of Fig.~\ref{fig:Dust_raw} shows the identical data shown in a log-linear style. However, any periodic undulation cannot be observed. Namely, the hierarchical structure (granulation) of tiny particles is necessary to induce periodic undulation during the columnar compression. At the same time, this result indicates that the stick-slip-like motion (sudden force drops) cannot be observed when the constituent particles are very small. In this situation, the compression behavior becomes similar to continuum case. Microscopically, deformation in this regime is probably dominated by rotation among particles. To observe the discrete nature of granular mechanics, large or hierarchically-structured particles are essential. When we use large rigid particles, friction-induced stick-slip motion governs the compression behavior. When large but fragile particles are used, collective breakage of particles causes the periodic undulation. The relation between microscopic deformation mode and macroscopic behavior is a key factor to be studied in future.

\bibliography{DustComp}

\end{document}